\documentclass[prd]{revtex4}
\usepackage{amsmath,bm}
\begin{document}
\title{The Plancherel Formula for the Universal Covering Group of
$\bm{SL(2,R)}$ Revisited}
\author{Debabrata Basu}
\affiliation{\\Saha Institute of Nuclear Physics,
1/AF, Bidhannagar, Kolkata - 700064, India}
\altaffiliation{Guest Scientist}
\date{\today}

\begin{abstract}
The Plancherel formula for the universal covering group of $SL(2, R)$ derived
earlier by Pukanszky on which Herb and Wolf build their Plancherel theorem for
general semisimple groups is reconsidered. It is shown that a set of unitarily
equivalent representations is treated by these authors as distinct. Identification
of this equivalence results in a Plancherel measure ($s\,\mathrm{Re}\tanh
\pi(s+\frac{i\tau}{2}),\ 0\leq\tau<1)$ which is different from the Pukanszky-Herb-Wolf
measure ($s\,\mathrm{Re}\tanh \pi(s+i\tau),\ 0\leq\tau<1)$.
\end{abstract}

\maketitle

\section{Introduction}

A major step ahead of the traditional concept of character was taken by Gel'fand and
Naimark\cite{img-man} in their definition in terms of the integral kernel of the group ring
which defines character as a linear functional on the group manifold. Let us denote
by $X$ the set of infinitely differentiable functions $x(g)$ on the group, which are
equal to zero outside a bounded set. If $g\rightarrow T_g$ is a representation of the group
$G$ we define the operator of the group ring as \begin{equation}T_x=\int d\mu(g)x(g)T_g \end{equation}where $d\mu(g)$ is the left and right invariant measure (assumed coincident) on $G$
and the integration extends over the entire group manifold. If we define
$$x_1x_2(g)=x(g)=\int x_1(g_1)x_2(g_1^{-1}g)d\mu(g_1)$$ then
\begin{equation}T_{x_1x_2}=T_{x_1}T_{x_2}\end{equation}
Let us suppose that $g\rightarrow T_g$ is a unitary representation of the group $G$ realized
in the Hilbert space $H$ of the functions $f(z)$ with the scalar product \begin{equation}(f,g)=\int\overline{f(z)}g(z)d\lambda(z) \end{equation}where $d\lambda(z)$ is the measure in $H$.
Then the operator $T_x$ is an integral operator with a kernel \begin{equation}T_x f(z)=\int
K(z,z_1)f(z_1)d\lambda(z_1) \end{equation}It then follows that $K(z,z_1)$ is a positive
definite Hilbert-Schmidt kernel satisfying
$$\int|K(z,z_1)|^2d\lambda(z)d\lambda(z_1)<\infty$$
Such a kernel has a trace \begin{equation}\mathrm{Tr}(T_x)=\int K(z,z)d\lambda(z)
\end{equation}Using the definition of the group ring $\mathrm{Tr}(T_x)$ can be
written in the form, \begin{equation}\mathrm{Tr}(T_x)=\int x(g)\pi(g)d\mu(g)
\end{equation}The function $\pi(g)$ is the character of the representation
$g\rightarrow T_g$. It should be noted that in this definition the matrix
representation of the group does not appear and it makes a complete synthesis of
the finite and infinite dimensional irreducible unitary representations.

The Vilenkin-Klimyk\cite{njav-auk} invariant method for the derivation of the Plancherel
formula consists in inverting the integral transform (6) to get $x(e)$. An important
step to achieve this objective is taken by evaluating the residue of the generalized
function $(x_3^2-x_2^2-x_1^2)^\lambda_+$ at the simple pole at $\lambda=-\frac{3}{2}$.

Our present framework is, however, more general than that of Vilenkin and Klimyk
insofar as it is not restricted to the integral and half integral representations of
Bargmann\cite{vb-1947}.

The irreducible unitary representations of the universal covering group of $SL(2,R)$
(denoted by $\widetilde{SL}(2, R))$ consists of four types:

\newcounter{bin1}\begin{list}{(\alph{bin1})}{\usecounter{bin1}}\item Principal series of representations $C_s^\epsilon$:
$$\begin{array}[t]{l} \displaystyle J_1^2+J_2^2-J_3^2=k(1-k);\hspace{.8em}k=\frac{1}{2}+is\\[2ex]
m=\epsilon\pm n,\ n=0,1,2,\ldots;\hspace{.8em}0\leq\epsilon<1\end{array}$$
\item Positive discrete series $D_k^+$:
$$k>0;\ m=k+n,\hspace{.8em} n=0,1,2,\ldots$$
\item Negative discrete series $D_k^-$:
$$k>0,\ m=-k-n,\hspace{.8em}n=0,1,2,\ldots$$
\item Exceptional series of representations $C_q^\epsilon$:
$$\begin{array}[t]{l}\displaystyle\epsilon(1-\epsilon)<q<\frac{1}{4};\hspace{.8em}0\leq\epsilon<1\\[2ex]
m=\epsilon\pm n,\hspace{.8em}n=0,1,2,\ldots\end{array}$$\end{list}

We omit the exceptional representations as they do not appear in the calculations.

The Plancherel formula for $\widetilde{SL}(2, R)$ was written down by Pukanszky\cite{lp}.
Following Pukanszky's work the more general problem of Plancherel theorem for
semisimple groups was attempted by Herb and Wolf\cite{rh-jaw} and by Duflo and Vergne\cite{md-mv} who
claim to be in agreement with Pukanszky. The edifice of Herb and Wolf's work on
general semisimple groups is built on Pukanszky's work on $\widetilde{SL}(2, R)$
(``We also need this special result to do the general case.'')\cite{rh-jaw}

However Pukanszky-Herb-Wolf formula for $\widetilde{SL}(2, R)$ suffers from a
serious flaw. All these authors treat a set of unitarily equivalent representations
as distinct. This flaw in Pukanszky's work has been carried into the later work of
Herb and Wolf\cite{rh-jaw}. As the rectification of the flaw for the general problem  is
likely to be more involved we start by investigating the problem for
$\widetilde{SL}(2, R)$.

We now assert that the representations $(\epsilon, s)$ and $(1-\epsilon, s)$
belonging to the principal series of $\widetilde{SL}(2, R)$ are unitarily
equivalent. The finite element of the group for the representation $(\epsilon, s)$
is given by (in terms of $SU(1, 1)$ parameters),
\begin{equation}
T_u^{(\epsilon, s)}f(z)=(\overline{\beta}\overline{z}+\alpha)^{-k-\epsilon}(\beta
z+\overline{\alpha})^{-k+\epsilon}f\left(\frac{\alpha z+\overline{\beta}}{\beta
z+\overline{\alpha}}\right)
\end{equation}
$$k=\frac{1}{2}+is,\ 0\leq\epsilon<1,\ |z|=1$$

The representations $C_s^{\epsilon}$ and $C_s^{1-\epsilon}$ are equivalent because
a fundamental property of the group ring ensures that
\begin{equation}
\mathrm{Tr}[T_x^{(1-\epsilon, s)}]=\mathrm{Tr}[T_x^{\epsilon, s}]\end{equation} It
has been shown in Sec.IB that the integral kernel of the ring is given by
\begin{equation}T_{x,\eta}^{(\epsilon, s)}g(\theta)=\int_0^{2\pi}K_{(\eta,
\epsilon)}(\theta,\theta_1)g(\theta_1)g(\theta_1)d\theta_1\end{equation} where
\begin{equation}f(z)=f(-ie^{i\theta})=g(\theta)\end{equation} is a single valued
function and \begin{equation}K_{(\eta, \epsilon)}^s(\theta,
\theta_1)=\frac{1}{4}\int
x(\underline{\theta}^{-1}\underline{k}\,\underline{\theta}_1)e^{i(\theta-\theta_1)\epsilon}|k_{22}|^{-1-2is}d\mu_l(k)\cos
2\pi\eta\epsilon\end{equation} where following the traditional procedure of
classical analysis we have defined the value of the functions on the real axis as
\begin{equation}f(x)=\frac{1}{2}[f(x+io)+f(x-io)]\end{equation} Thus for $C_s^{1-\epsilon}$ we have
\begin{equation}K_{(\eta, 1-\epsilon)}^s(\theta, \theta_1)=\frac{1}{4}\int
x(\underline{\theta}^{-1}\underline{k}\,\underline{\theta}_1)e^{i(\theta-\theta_1)(1-\epsilon)}
|k_{22}|^{-1-2is}d\mu_l(k)\cos 2\pi\eta\epsilon\end{equation}

Thus although the integral kernels differ the $\epsilon$ dependent term from the
exponential drops out from the traces and we have$$\begin{array}[t]{ll}
\mathrm{Tr}\left[T_{x,\eta}^{(\epsilon, s)}\right]&=\displaystyle\int K_{(\eta,
\epsilon)}^s(\theta, \theta)d\theta\\[2ex]
&=\displaystyle\int K_{(\eta,
1-\epsilon)}(\theta,\theta)d\theta=\mathrm{Tr}\left[T_{x,\eta}^{(1-\epsilon,
s)}\right]\end{array}$$

We may, therefore, write\setcounter{equation}{13} \begin{eqnarray}
\mathrm{Tr}[T_x^{(\epsilon,
s)}]&=&\theta(2\epsilon)\theta(1-2\epsilon)\mathrm{Tr}\left[T_x^{(\epsilon,
s)}\right]
+\theta(2-2\epsilon)\theta(2\epsilon-1)\mathrm{Tr}\left[T_x^{\epsilon, s}\right]\nonumber\\[1ex]
&=&\theta(2\epsilon)\theta(1-2\epsilon)\mathrm{Tr}\left[T_x^{(\epsilon,
s)}\right]+\theta(2-2\epsilon)\theta(2\epsilon-1)\mathrm{Tr}\left[T_x^{(1-\epsilon, s)}\right]\end{eqnarray}

This point has not been taken care of by Pukanszky or by Herb and Wolf.

The problem consists of two parts (1) evaluation of the character of the
representations of type (a), (b) and (c); (2)~the inversion problem and the
subsequent computation of $x(e)$. We now announce the Plancherel formulas due to (i)
Pukanszky, Herb and Wolf (in our notation) (ii) the present author \begin{tabbing}\indent(i) $\displaystyle
x(e)=\frac{2}{\pi^2}\int_0^\infty ds\int_0^1 d\tau s\,
\mathrm{Re}\tanh\pi(s+i\tau)\mathrm{Tr}[T_x^{(\tau, s)}]
+\frac{2}{\pi^2}\int_{\frac{1}{2}}^\infty
dk\left(k-\frac{1}{2}\right)\mathrm{Tr}\left[T_x^{k+}+T_x^{k-}\right]$\`(15a)\\[1ex]
\`(Pukanszky-Herb-Wolf)\\[1ex]
\indent(ii) $\displaystyle x(e)=\frac{2}{\pi^2}\int_0^\infty ds\int_0^1 d\tau s\,
\mathrm{Re}\tanh\pi\left(s+\frac{i\tau}{2}\right)\mathrm{Tr}[T_x^{\left(\frac{\tau}{2},
s\right)}]+\frac{2}{\pi^2}\int_{\frac{1}{2}}^\infty
dk\left(k-\frac{1}{2}\right)\mathrm{Tr}[T_x^{k+}+T_x^{k-}]$\`(15b)\\
\`(author)\end{tabbing}

\section{Evaluation of Character}

\subsection{Elliptic and Hyperbolic elements of the group}

The group $SU(1,1)$ consists of pseudounitary unimodular matrices
$$u=\left(\begin{array}{cc}\alpha&\beta\\[.5ex]
\overline{\beta}&\overline{\alpha}\end{array}\right),\ |\alpha|^2-|\beta|^2=1$$ and is isomorphic to the
real unimodular matrices
$$g=\left(\begin{array}{cc}a&b\\[.5ex]
c&d\end{array}\right),\ ad-bc=1.$$

\noindent A particular choice of the isomorphism kernel is
$$\sigma=\frac{1}{\sqrt{2}}\left(\begin{array}{cc}1&i\\[.2ex]
i&1 \end{array}\right)$$
so that $u=\sigma g\sigma^{-1}$. Thus
$$\alpha=\frac{1}{2}[(a+d)-i(b-c)],\ \beta=\frac{1}{2}[(b+c)-i(a-d)]$$

The elements of the group $SU(1,1)$ or $(SL(2,R))$ may be divided into three
subsets:

(i) elliptic (ii) hyperbolic (iii) parabolic.

\noindent We define them as follows. Let $$\alpha=\alpha_1+i\alpha_2;\
\beta=\beta_1+i\beta_2\vspace{-2pt}$$ so that\vspace{-2pt} $$
|\alpha|^2-|\beta|^2=\alpha_1^2+\alpha_2^2-\beta_1^2-\beta_2^2=1$$ The elliptic
elements are those for which $$\alpha_2^2-\beta_1^2-\beta_2^2>0$$ Hence if we
set $\alpha_2'=\sqrt{\alpha_2^2-\beta_1^2-\beta_2^2}$
we have $\alpha_1^2+\alpha_2'^{\,2}=1$, i.e.\ $-1<\alpha_1<1$. On the other
hand the hyperbolic elements are those for which
$$\alpha_2^2-\beta_1^2-\beta_2^2<0$$
Hence if we write $\alpha_2'=\sqrt{\beta_1^2+\beta_2^2-\alpha_2^2}$
we have $\alpha_1^2-\alpha_2'^{\,2}=1$ so that $|\alpha_1|>1$. We exclude the
parabolic class corresponding to
$$\alpha_2=\sqrt{\beta_1^2+\beta_2^2}$$
as this is a submanifold of lower dimensions. In a previous paper\cite{db-1997} we have shown
that the elliptic elements can be decomposed as
$$u=\epsilon(\eta)a(\rho)\epsilon(\theta_0)a^{-1}(\rho)\epsilon^{-1}(\eta)$$
$0<\theta_0<2\pi,\ 0\leq\eta\leq4\pi,\ 0\leq\rho<\infty$, where
$$\epsilon(\theta_0)=\left(\begin{array}{cc}e^{i\theta_0/2}&0\\[.5ex]
0&e^{-i\theta_0/2}\end{array}\right)$$
$$a(\rho)=\left(\begin{array}{cc} \cosh\frac{\rho}{2}&\sinh\frac{\rho}{2}\\[1.5ex]
\sinh\frac{\rho}{2}&\cosh\frac{\rho}{2}\end{array}\right)$$
Thus$$\alpha=\cos\frac{\theta_0}{2}+i\sin\frac{\theta_0}{2}\cosh\rho$$
$$\beta=-ie^{i\eta}\sin\frac{\theta_0}{2}\sinh\rho$$
The corresponding invariant measure is given by \def\theequation{16a}\begin{equation}d\mu(u)=\sin^2\frac{\theta_0}{2}\,\frac{d\theta_0}{2}\sinh\rho\,d\rho\,d\eta \end{equation}On the
other hand writing the eigenvalues as $\lambda=\mathrm{sgn}\lambda
e^{\pm\mathrm{sgn}\lambda\frac{\sigma}{2}}$, the hyperbolic elements can be parameterized
as\cite{db-1997}\def\theequation{16b} \begin{equation}\begin{array}{rl}
\alpha&\displaystyle=\mathrm{sgn}\lambda\cosh\frac{\sigma}{2}+i\sinh\frac{\sigma}{2}\sinh\rho\\[2ex]
\beta&\displaystyle=-ie^{-i\theta}\sinh\frac{\sigma}{2}\cosh\rho \end{array}\end{equation}$0\leq\theta\leq4\pi,\ 0\leq\sigma<\infty,\ 0\leq\rho<\infty$.
 The invariant measure for the hyperbolic case is given by \def\theequation{16c}\begin{equation}d\mu(u)=\sinh^2\frac{\sigma}{2}\,\frac{d\sigma}{2}\cosh\rho\,d\rho\,d\theta \end{equation}
\subsection{The character of the principal series of representation}

The principal series is realized in the Hilbert space of functions defined on the
unit circle $|z|=1$. The finite element of the group is given
by,\setcounter{equation}{16}\def\theequation{\arabic{equation}}
\begin{equation}T_u^{(\epsilon,s)}f(z)=(\beta
z+\overline{\alpha})^{-k+\epsilon}(\overline{\beta}\,\overline{z}+\alpha)^{-k-\epsilon}f\left(\frac{\alpha
z+\overline{\beta}}{\beta z+\overline{\alpha}}\right) \end{equation}with
$\displaystyle k=\frac{1}{2}+is,\ 0\leq\epsilon<1$. These representations are
unitary with respect to the scalar product
\begin{equation}(f,g)=\int\overline{f(z)}g(z)d\theta,\ z=e^{i\theta}
\end{equation}It will be shown presently that the representation $\epsilon$ and $1-\epsilon$ are
unitarily equivalent. Thus the representations $0\leq\epsilon\leq\frac{1}{2}$ and
$\frac{1}{2}\leq\epsilon<1$ are equivalent.

We now construct the operator of the group ring \begin{equation}T_x^{(\epsilon,s)}=\int
d\mu(u)x(u)T_u^{(\epsilon,s)} \end{equation}where $x(u)$ is an arbitrary test function on the
group which vanishes outside a bounded set. If we define
$$x^\dagger(u)=\overline{x(u^{-1})}$$
we have from Eq.\ (2),
\begin{equation}
x_1^\dagger x_2(e)=\int\overline{x_1(u^{-1})}x_2(u^{-1})d\mu(u)
=\int\overline{x_1(u)}x_2(u)d\mu(u)\end{equation}

The operator of the group ring is now given by \setcounter{equation}{20}\begin{equation}T_x^{(\epsilon,s)}f(z)=\int d\mu(u)x(u)(\beta
z+\overline{\alpha})^{-k+\epsilon}(\overline{\beta}\,\overline{z}+\alpha)^{-k-\epsilon}f\left(\frac{\alpha
z+\overline{\beta}}{\beta z+\overline{\alpha}}\right) \end{equation}Setting $z=-ie^{i\theta}$ and
performing the left translation, \begin{equation}u\rightarrow\underline{\theta}^{-1}u,\ \underline{\theta}=\left(\begin{array}{cc}e^{i\theta/2}&0\\[1ex]
0&e^{-i\theta/2}\end{array}\right)\end{equation}we have \begin{equation}T_xf(-ie^{i\theta})=\int
d\mu(u)x(\theta^{-1}u)e^{i\theta\epsilon}(-i\beta+\overline{\alpha})^{-k+\epsilon}(i\overline{\beta}+\alpha)^{-k-\epsilon}f\left(\frac{-i\alpha+\overline{\beta}}{-i\beta+\overline{\alpha}}\right)
\end{equation}We now perform the Iwasawa decomposition of $SL(2,R)$ \begin{equation}g=k\theta_1,\
k=\left(\begin{array}{cc} k_{11}&k_{12}\\[.5ex]
0&k_{22} \end{array}\right),\ k_{11}k_{22}=1,\
\theta_1=\left(\begin{array}{cc}\cos\frac{\theta_1}{2}&-\sin\frac{\theta_1}{2}\\[.5ex]
\sin\frac{\theta_1}{2}&\cos\frac{\theta_1}{2}\end{array}\right) \end{equation}so that
$$u=\underline{k}\,\underline{\theta_1},\ \underline{k}=\sigma k\sigma^{-1}\ \mbox{etc.}$$
We now write \begin{eqnarray}(i\overline{\beta}+\alpha)&=&(k_{22}+i0)e^{i\theta_1/2}\\[1ex]
(-i\beta+\overline{\alpha})&=&(k_{22}-i0)e^{-i\theta_1/2} \end{eqnarray} and set \setcounter{equation}{26}\begin{equation}g(\theta)=f(-ie^{i\theta}) \end{equation}Then we have \begin{equation}T_{x,\eta}^{(\epsilon,s)}g(\theta)=\int_0^{2\pi}K_\eta(\theta,\theta_1)g(\theta_1)d\theta_1
\end{equation}with \begin{equation}K_\eta(\theta,\theta_1)=\frac{1}{4}\int
x(\underline{\theta}^{-1}\,\underline{k}\,\underline{\theta}_1)e^{i(\theta-\theta_1)\epsilon}|k_{22}|^{-2k}\times\cos2\pi\eta\epsilon\,d\mu_l(k)
\end{equation}In obtaining the integral kernel $K_\eta(\theta,\theta_1)$ of the group ring we
have written \begin{equation}\begin{array}{rl}\multicolumn{2}{c}{(k_{22}\pm i0)=|k_{22}|e^{\pm
i\pi\eta}}\\[1ex]
\eta&=0,\ \mbox{for}\ k_{22}>0\\[1ex]
\eta&=1,\ \mbox{for}\ k_{22}<0\end{array}\end{equation}and following the traditional procedure of
classical analysis we have defined the value of the function on the real axis as,
\begin{equation}f(x)=\frac{1}{2}[f(x+i0)+f(x-i0)]\end{equation}Since the kernel is of the Hilbert-Schmidt
type
$$\mathrm{Tr}(T_{x,\eta}^{(\epsilon,s)})=\int_0^{2\pi}K_\eta(\theta,\theta)d\theta$$
which can be written in the form, \begin{equation}\mathrm{Tr}(T_{x,\eta}^{(\epsilon,s)})=\frac{1}{4}\int_\Theta d\theta\int
d\mu_l(k)|k_{22}|^{-2k}x(\underline{\theta}^{-1}\,\underline{k}\,\underline{\theta})\cos2\pi\epsilon\eta
\end{equation}Before proceeding any further we note that
$\underline{\theta}^{-1}\,\underline{k}\,\underline{\theta}$ represents a hyperbolic element of
$SU(1,1)$: \begin{equation}u=\underline{\theta}^{-1}\,\underline{k}\,\underline{\theta} \end{equation}Calculating the trace of
the $2\times 2$ matrix
$$2\alpha_1=k_{22}+\frac{1}{k_{22}}$$
The above equation for the elliptic element yields
$$k_{22}^2-2k_{22}\cos\frac{\theta_0}{2}+1=0$$
which has no real solution. Thus the elliptic elements do not contribute to the
character of the principal series of representations.

Following ref.\ 8 we may now show that every hyperbolic element of $SU(1,1)$ can be
represented as
$$u=\underline{\theta}^{-1}\,\underline{k}\,\underline{\theta}$$

Here $k_{11}=\lambda^{-1},\ k_{22}=\lambda$ are the eigenvalues of the matrix $u$
taken in any order. It follows that for a given choice of $\lambda$, the parameters
$\theta$ and $k_{12}$ are uniquely determined. We note that there are exactly two
representations of the matrix $g=\sigma^{-1}u\sigma\in SL(2,R),\
\sigma=\frac{1}{\sqrt2}\left(\begin{array}{cc}1&i\\[.5ex]
i&1\end{array}\right)$ by means of the formula (33) corresponding to two distinct
possibilities
$$\begin{array}{ll}|k_{11}|=|\lambda|^{-1}=e^{\sigma/2}&\hspace{2em}
|k_{22}|=|\lambda|=e^{-\sigma/2}\\[1ex]
|k_{11}|=|\lambda|^{-1}=e^{-\sigma/2}&\hspace{2em}|k_{22}|=|\lambda|=e^{\sigma/2}\end{array}$$

Following Gelfand and Naimark\cite{img-man} (in $SL(2,C))$ let us now remove from $K$ the
elements with $k_{11}=k_{22}=1$. This operation cuts the group $K$ into two
connected disjoint components. In view of this partition the integral in (32) is
represented in the form of the sum of two integrals
\setcounter{equation}{33}\begin{equation}
\mathrm{Tr}(T_{x,\eta}^{(\epsilon,s)})=\frac{1}{4}\int_\Theta
d\theta\int_{K_1}d\mu_l(k)|k_{22}|^{-2k}x(\underline{\theta}^{-1}\,
\underline{k}\,\underline{\theta})\cos2\pi\epsilon\eta+\frac{1}{4}\int_\Theta
d\theta\int_{K_2}d\mu_l(k)|k_{22}|^{-2k}x(\underline{\theta}^{-1}\,\underline{k}\,\underline{\theta})\cos2\pi\epsilon\eta\end{equation}

As $\theta$ runs over the subgroup $\Theta$ and $k$ runs over the components $K_1$
or $K_2$ the matrix $u=\underline{\theta}^{-1}\,\underline{k}\,\underline{\theta}$ runs over the
hyperbolic elements of $SU(1,1)$ (or equivalently $SL(2,R)$). Following ref.\ 8 it
can now be proved that in $K_1$ or $K_2$, \begin{equation}d\mu_l(k)d\theta=\frac{4|k_{22}|d\mu(u)}{|k_{11}-k_{22}|}\end{equation}so that \begin{equation}\mathrm{Tr}(T_x^{(\epsilon,s)})=\sum_{\eta=0,1}\int
x(u)\pi^{(\epsilon,s)}_\eta(u)d\mu(u)\end{equation}where \begin{equation}\pi^{(\epsilon,s)}_\eta(u)=\frac{\cos
s\sigma}{\sinh\frac{\sigma}{2}}\cos2\pi\epsilon\eta \end{equation}
\subsection{The positive discrete series $\bm{D^+_k}$}

The finite element of the group for the representation is given by, \begin{equation}T^{k+}_uf(z)=(\beta z+\overline{\alpha})^{-2k}f\left(\frac{\alpha z+\overline{\beta}}{\beta
z+\overline{\alpha}}\right)\end{equation}where $f(z)$ is an analytic function regular within the
unit disc. This representation is unitary with respect to the scalar product \begin{equation}(f,g)=\int_{|z|<1}\overline{f(z)}g(z)d\lambda(z)\end{equation}where \begin{equation}d\lambda(z)=\frac{2k-1}{\pi}(1-|z|^2)^{2k-2} \end{equation}\pagebreak

The integral converges in the usual sense for $k>\frac{1}{2}$. For $0<k<\frac{1}{2}$ the
integral is to be understood in the sense of its regularization (analytic
continuation). Thus \begin{tabbing}\hspace{.8in}\=\kill \>$\displaystyle(f,g)=\frac{2k-1}{2\pi(1-e^{4\pi
ik})}\int_{\Sigma}dt(1-t)^{2k-2}\times\int\overline{f(z)}g(z)d\theta$\`(41a)\\[1.5ex]
\>$z=\sqrt te^{i\theta}$\`(41b)\end{tabbing} where $\sum$ is a contour (in the $t$ plane) that
starts from the origin along the positive real axis, encircles the point $+1$
counter-clockwise and returns to the origin along the positive real axis.

The principal vector\cite{vg-1961,ies} in this Hilbert space is given by\cite{db-1997}, \setcounter{equation}{41}\begin{equation}e_z(z_1)=(1-\overline{z}z_1)^{-2k}\end{equation}so that \begin{equation}f(z)=(e_z,f)=\int_{|z_1|<1}(1-z\overline{z}_1)^{-2k}f(z_1)d\lambda(z_1)\end{equation}
The action of the group ring \begin{equation}T^{k+}_x=\int x(u)T^{k+}_ud\mu(u)\end{equation}where
$d\mu(u)$ is the invariant measure on $SU(1,1)$ is given by, \begin{equation}T^{k+}_xf(z)=\int
x(u)(\beta z+\overline{\alpha})^{-2k}\times f\left(\frac{\alpha z+\overline{\beta}}{\beta
z+\overline{\alpha}}\right)d\mu(u)\end{equation}Now we use Eq.\ (43) to write \begin{equation}f\left(\frac{\alpha
z+\overline{\beta}}{\beta z+\overline{\alpha}}\right)=\int_{|z_1|<1}\left[1-\frac{(\alpha
z+\overline{\beta})\overline{z}_1}{(\beta z+\overline{\alpha})}\right]^{-2k}f(z_1)d\lambda(z_1)\end{equation}This immediately yields \begin{equation}T^{k+}_xf(z)=\int_{|z_1|<1}K(z,z_1)f(z_1)d\lambda(z_1)\end{equation}where \begin{equation}K(z,z_1)=\int(\beta z+\overline{\alpha})^{-2k}\left[1-\frac{(\alpha
z+\overline{\beta})\overline{z}_1}{(\beta z+\overline{\alpha})}\right]^{-2k}x(u)d\mu(u) \end{equation}
Since the kernel is again of the Hilbert Schmidt type we have \begin{equation}\mathrm{Tr}(T^{k+}_x)=\int_{|z|<1}K(z,z)d\lambda(z)\end{equation}which can be written in the form,
\begin{equation}\mathrm{Tr}(T_x^{k+})=\int d\mu(u)x(u)\pi^{k+}(u)\end{equation}where
\begin{equation}\pi^{k+}(u)=\int_{|z|<1}d\lambda(z)[(\beta z+\overline{\alpha})-(\alpha
z+\overline{\beta})\overline{z}]^{-2k}\end{equation}We now substitute \begin{equation}z=\tanh\frac{\tau}{2}e^{i\phi},\
0\leq\tau<\infty;\ 0\leq\phi\leq2\pi\end{equation}Thus, as shown in I, for the elliptic
elements (see Eq.\ (13)). \begin{equation}\pi^{k+}=\frac{2k-1}{4\pi}\int_{\tau=0}^\infty\int^{2\pi}_{\phi=0}\left[\cos\frac{\theta_0}{2}-i\sin\frac{\theta_0}{2}\hat{n}\cdot\hat{r}\right]^{-2k}\sinh\tau
d\phi d\tau\end{equation}where $\hat{n}$ and $\hat{r}$ are unit time-like vectors,
\begin{equation}\begin{array}{rl}\hat{n}&=(\cosh\rho,-\sinh\rho\sin\eta,\sinh\rho\cos\eta)\\[1ex]
\hat{r}&=(\cosh\tau,\sinh\tau\sin\phi,\sinh\tau\cos\phi)\end{array}\end{equation}and
$\hat{n}\cdot\hat{r}$ is the Lorentz invariant form,
$$\hat{n}\cdot\hat{r}=\hat{n}_3\hat{r}_3-\hat{n}_2\hat{r}_2-\hat{n_1}\hat{r}_1$$

We now perform a Lorentz transformation such that the time axis coincides with the
time like vector $\hat{n}$. Thus
$$\hat{n}\cdot\hat{r}=\cosh\tau$$ and we have \begin{equation}\pi^{k+}=\frac{2k-1}{2}\int_0^\infty
d\tau\sinh\tau\left[\cos\frac{\theta_0}{2}-i\sin\frac{\theta_0}{2}\cosh\tau\right]^{-2k}
\end{equation}The integration is quite elementary and we have\cite{img-mig-iip}
\begin{equation}\pi^{k+}(u)=\frac{e^{\frac{i\theta_0}{2}(2k-1)}}{e^{-\frac{i\theta_0}{2}}-e^{\frac{i\theta_0}{2}}}
\end{equation}
For the hyperbolic elements we substitute Eq.\ (16b) and perform a Lorentz
transformation ($\hat{n}$ is now a space-like vector) such that the first space axis
concides with $\hat{n}$. Thus
\begin{equation}\pi^{k+}(u)=\frac{2k-1}{4\pi}\int^\infty_0d\tau\sinh\tau\int^{2\pi}_0d\phi\left[\mathrm{sgn}\lambda\cosh\frac{\sigma}{2}-i\sinh\frac{\sigma}{2}\sinh\tau\cos\phi\right]^{-2k}
\end{equation}
Evaluation of this integral can be carried out as in ref.\ 8 and using the
definition (31) we have
\begin{equation}\pi^{k+}(u)=\frac{e^{-\frac{\sigma}{2}(2k-1)}}{e^{\frac{\sigma}{2}}-e^{-\frac{\sigma}{2}}}\cos2\pi\epsilon'\eta,\
\eta=0,1\end{equation}\vspace{-15pt}$$\begin{array}{rlll}k&=n+1-\epsilon'&\mbox{for}&\displaystyle 0\leq\epsilon'\leq\frac{1}{2}\\[1ex]
&=n+2-\epsilon'&&\displaystyle\frac{1}{2}\leq\epsilon'<1\end{array}$$

\subsection{Negative discrete series $\bm{D_k^-}$}

The finite element of the group in this case is taken to be \begin{equation}T^{k-}_uf(z)=(\alpha+\overline{\beta}z)^{-2k}f\left(\frac{\beta+\overline{\alpha}z}{\alpha+\overline{\beta}z}\right)
\end{equation}where $f(z)$ is analytic within the unit disc. The scalar product with respect
which $T^{k-}_u$ is unitary is given by
\begin{equation}\begin{array}{rl}(f,g)&\hspace{-.65em}\displaystyle=\int\overline{f(z)}g(z)d\lambda(z)\\[2.5ex]
d\lambda(z)&\hspace{-.65em}\displaystyle=\frac{2k-1}{\pi}(1-|z|^2)^{2k-2}\end{array}\end{equation}The principal vector is as
before
$$e_z(z_1)=(1-\overline{z}z_1)^{-2k}$$ so that \begin{equation}f(z)=\int_{|z_1|<1}(1-z\overline{z}_1)^{-2k}f(z_1)d\lambda(z_1)\end{equation}\pagebreak

\noindent Proceeding in the same way as before
\begin{equation}\pi^{k-}(u)=\int_{|z|<1}[(\alpha+\overline{\beta}z)-(\beta+\overline{\alpha}z)\overline{z}]^{-2k}d\lambda(z)
\end{equation}Setting $z=\tanh\frac{\tau}{2}e^{i\phi}$ we have as before, for the elliptic
elements \begin{equation}\pi^{k-}(u)=\frac{2k-1}{2}\int
d\tau\sinh\tau\left[\cos\frac{\theta_0}{2}+i\sin\frac{\theta_0}{2}\cosh\tau\right]^{-2k}=\frac{e^{-\frac{i\theta_0}{2}(2k-1)}}{e^{\frac{i\theta_0}{2}}-e^{-\frac{i\theta_0}{2}}}\end{equation}For the hyperbolic elements a parallel calculation gives \begin{equation}\pi^{k-}(u)=\frac{e^{-\frac{\sigma}{2}(2k-1)}\cos2\pi\epsilon'\eta}{[e^{\frac{\sigma}{2}}-e^{-\frac{\sigma}{2}}]},\
\eta=0,1\end{equation}and $\epsilon'$ is defined in the same way as before.

\section{The problem of inversion and the Plancherel formula}

Let us start from, \setcounter{equation}{64}\begin{equation}\mathrm{Tr}(T_x^{(\epsilon,s)})=\sum_{\eta=0,1}\left[\int_{\mathrm{elliplic}}x(u)\pi^{(\epsilon,s)}(u)d\mu(u)+\int_{\mathrm{hyperbolic}}x(u)\pi^{(\epsilon,s)}_\eta(u)d\mu(u)\right]
\end{equation}

For the principal series (as indicated by the index $s$) the first term is zero. For
the hyperbolic elements \begin{tabbing}\hspace{5.5cm}\=\kill \>$\displaystyle
d\mu(u)=\sinh^2\frac{\sigma}{2}\,\frac{d\sigma}{2}\cosh\rho d\rho
d\theta$\`(66a)\\[2ex]
\>$\displaystyle\pi^{(\epsilon,s)}_\eta(u)=\frac{\cos
s\sigma}{\sinh\frac{\sigma}{2}}\cos2\pi\epsilon\eta$\`(66b)\end{tabbing} Thus if we define
$$\phi_\eta(t)=\int x(\theta,\rho;\eta,t)\cosh\rho d\rho d\theta$$
we have $\displaystyle\frac{\sigma}{2}=t,\hspace{1em}\eta=0,1$\setcounter{equation}{66}\begin{equation}\mathrm{Tr}(T^{(\epsilon,s)}_x)=\int^\infty_0[\phi_0(t)+\cos2\pi\epsilon\phi_1(t)]\sinh
t\cos2stdt\end{equation}Now we shall divide $\epsilon$ into two regions,
$0\leq\epsilon\leq\frac{1}{2}$, and $\frac{1}{2}<\epsilon<1$. Thus
$$\mathrm{Tr}(T_x^{\epsilon,s})=\theta(2\epsilon)\theta(1-2\epsilon)\mathrm{Tr}(T_x^{(\epsilon,s)})+\theta(2\epsilon-1)\theta(2-2\epsilon)\mathrm{Tr}(T^{(\epsilon,s)}_x)$$ These two regions yield two sets of representations
which are unitarily equivalent.

We now have $$\mathrm{Tr}(T^{k+}_x+T^{k-}_x)=\int[\phi_0(t)+\cos2\pi\epsilon'\phi_1(t)]\times\sinh tdt e^{-\nu t}-\int^\pi_0d\theta\sin^2\theta\int\int
x(\eta,\rho;\theta)\frac{\sin\nu\theta}{\sin\theta}\times\sinh\rho d\rho d\eta$$
where $\nu=\nu(\epsilon')=2k-1$

We now set \begin{equation}\int\int x(\eta,\rho;\theta)\sinh\rho d\rho d\eta=F(\theta)\end{equation}so that
\begin{equation}\int^\pi_0F(\theta)\sin\theta\sin\nu\theta
d\theta=\int^\infty_0[\phi_0(t)+\cos2\pi\epsilon'\phi_1(t)]\sinh te^{-\nu
t}dt-\mathrm{Tr}(T^{k+}_x+T^{k-}_x)\end{equation}
Applying the inversion formula for the Fourier cosine transform
\begin{equation}
[\phi_0(t)+\cos2\pi\epsilon'\phi_1(t)]\sinh
t=\frac{4}{\pi}\int^\infty_0ds\cos2st[\theta(1-2\epsilon')\mathrm{Tr}(T^{(\epsilon',s)}_x)\theta(2\epsilon')
+\theta(2\epsilon'-1)\theta(2-2\epsilon')\mathrm{Tr}(T^{(\epsilon',s)}_x)]\end{equation}
we obtain\begin{tabbing}\hspace{2em}\=\kill
\>$\displaystyle\int F(\theta)\sin\theta\sin\nu\theta d\theta=\frac{4}{\pi}\int ds\int
dt\,\mathrm{Tr}(T^{(\epsilon',s)}_x)\theta(1-2\epsilon')e^{-\nu t}\cos2st\theta(2\epsilon')$\\[1.5ex]
\`${}+\displaystyle\frac{4}{\pi}\int ds\int
dt\,\mathrm{Tr}(T^{(1-\epsilon',s)}_x)\theta(2-2\epsilon')\theta(2\epsilon'-1)e^{-\nu
t}\cos2st-\mathrm{Tr}(T^{k+}_x+T^{k-}_x)$\hspace{1em}(71)\\[1ex] The above formula implies that in
the Fourier sine transform of the function $F(\theta)\sin\theta$,\\[1ex]
\>$\displaystyle F(\theta)\sin\theta=\frac{2}{\pi}\int^\infty_0c(\nu)\sin\nu\theta
d\nu$\\[2ex]
\>$\displaystyle c(\nu)=\frac{4}{\pi}\int ds\int dt\,\mathrm{Tr}(T^{(\epsilon,s)}_x)e^{-\nu
t}\theta(1-2\epsilon')\cos2st\theta(2\epsilon')$\\[1.5ex]
\`${}+\displaystyle\frac{4}{\pi}\int ds\int dt\,\mathrm{Tr}(T^{(1-\epsilon',s)}_x)e^{-\nu
t}\theta(2-2\epsilon')\theta(2\epsilon'-1)\cos2st-\mathrm{Tr}(T^{k+}_x+T^{k-}_x)$\hspace{2em}(72)\\[1ex]
Thus we have\\[1ex]
\>$F(\theta)\sin\theta$ \=$\displaystyle=\frac{8}{\pi^2}\int ds\int
dt\theta(1-2\epsilon')\theta(2\epsilon')\mathrm{Tr}(T^{(\epsilon',s)}_x)\int e^{-\nu
t}\cos2st
\sin\nu\theta d\nu$\\[1ex]
\>\>\hspace{5em}$\displaystyle{}+\frac{8}{\pi^2}\int ds\int
dt\theta(2-2\epsilon')\theta(2\epsilon'-1)\mathrm{Tr}(T^{(1-\epsilon',s)}_x)\cos2st\int
e^{-\nu t}\sin\nu\theta d\nu$\\[1ex]
\`$\displaystyle{}-\int^\infty_0\mathrm{Tr}(T^{k+}_x+T^{k-}_x)\sin\nu\theta d\nu$\end{tabbing}
In the
first integral, $$k=n+1-\epsilon'\hspace{1em}\nu=2k-1=2n+1-\tau\hspace{1em}d\nu=-d\tau$$ Similarly
in the second $$k=n+2-\epsilon'\hspace{1em}\nu=2n+1+\tau\hspace{1em}\tau=2-2\epsilon';\hspace{1em}d\nu=d\tau$$ Thus we have \begin{tabbing}\hspace{.3in}\=\kill \>$F(\theta)\sin\theta$
\=$\displaystyle=\frac{8}{\pi^2}\int^\infty_0ds\int^1_0d\tau
\mathrm{Tr}(T_x^{\left(\frac{\tau}{2},s\right)})\int^\infty_0dt\cos 2st$\\[1.5ex]
\>\>\hspace{2em}\hspace{2em}${}\times\displaystyle\sum^\infty_{n=0}[e^{-(2n+1-\tau)t}\sin(2n+1-\tau)\theta+e^{-(2n+1+\tau)t}\sin(2n+1+\tau)\theta]$\\[.5ex]
\`${}-\displaystyle\frac{2}{\pi}\int \mathrm{Tr}(T^{k+}_x+T^{k-}_x)\sin\nu\theta
d\nu$\hspace{2em}(73)\\[1ex]
The summation can be easily carried out recalling that\\[1.5ex]
\>$\displaystyle\sum
e^{-(2n+1\mp\tau)t}\sin(2n+1\mp\tau)\theta=\mathrm{Im}e^{-t(1\mp\tau)+i(1\mp\tau)\theta}
\times\displaystyle\sum e^{-2nt+2in\theta}$\`(74)\\[1.5ex]
and using the standard summation formula for the geometric series.
Thus\\[1ex]
\>$\displaystyle F(\theta)\sin\theta=\frac{8}{\pi^2}\int^\infty_0ds\int^1_0d\tau
\mathrm{Tr}(T^{\left(\frac{\tau}{2},s\right)}_x)\times\left[\sin(1+\tau)\theta\int^\infty_0dt\frac{\cos2st\cosh(1-\tau)t}{\cosh2t-\cos2\theta}\right.$\\[2.5ex]
\`${}+\displaystyle\left.\sin(1-\tau)\theta\int^\infty_0dt\frac{\cos2st\cosh(1+\tau)t}{\cosh2t-\cos2\theta}\right]
-\displaystyle\frac{2}{\pi}\int^\infty_0d\nu\sin\nu\theta\mathrm{Tr}(T^{k+}_x+T^{k-}_x)$\hspace{1.5em}(75) \end{tabbing}

The integrals can be written in the form,
$$\frac{1}{2}\int^\infty_{-\infty}e^{2ist}\frac{\cosh(1\mp\tau)tdt}{\cosh2t-\cos2\theta}$$
which can be evaluated by the sum of the residues at the poles in the upper
half-plane located at $$t=i(\theta+n\pi),\ n=0,1,2,\ldots$$ and at
$$t=i(n\pi-\theta),\ n=1,2,3\ldots$$ It can be easily ascertained that
$$\sin(1+\tau)\theta\int^\infty_0\frac{\cos2st\cosh(1-\tau)t}{\cosh2t-\cos2\theta}dt+\sin(1-\tau)\theta\int^\infty_0\frac{\cos2st\cosh(1+\tau)t}{\cosh2t-\cos2\theta}dt
=\frac{\pi}{2}\mathrm{Re}\frac{\cosh\left[\pi\left(s+\frac{i\tau}{2}\right)-2s\theta\right]}{\cosh\pi\left(s+\frac{i\tau}{2}\right)}$$
We therefore obtain
\setcounter{equation}{75}\begin{equation}F(\theta)\sin\theta=\frac{4}{\pi}\int^\infty_0ds\int^1_0d\tau
\mathrm{Tr}(T^{\left(\frac{\tau}{2},s\right)}_x)\mathrm{Re}\frac{\cosh\left[\pi\left(s+\frac{i\tau}{2}\right)-2s\theta\right]}{\cosh\pi\left(s+\frac{i\tau}{2}\right)}
-\frac{4}{\pi}\int dk\sin(2k-1)\theta \mathrm{Tr}[T^{k+}_x+T^{k-}_x]\end{equation}

(b) It now remains to relate $x(e)$ with $F(\theta)\sin\theta$. For this we equate
two different calculations for the residue at the pole at $\lambda=-\frac{3}{2}$ of the
generalized function
$$(x^2_3-x^2_2-x^2_1)^\lambda_+$$ as an analytic function of $\lambda$.
Let us consider \begin{equation}I(\lambda)=\int_{x_0>0}(x^2_3-x^2_2-x^2_1)^\lambda_+x(u)d\mu(u)\end{equation}where $d\mu(u)$
stands for the invariant measure for the elliptic elements (as given by Eq. (14))
and
$$x_3=\sin\theta\cosh\rho\hspace{1em}x_2=\sin\theta\sinh\rho\sin\eta$$
$$x_1=\sin\theta\sinh\rho\cos\eta;\ 0\leq\eta\leq4\pi,\ 0\leq\rho<\infty;\
0\leq\theta\leq\pi$$ where $\displaystyle\theta=\frac{\theta_0}{2},\ \rho,\eta$ are defined by
Eq.\ (13) and $x_0=\cos\theta$.

\noindent It now immediately follows \begin{equation}d\mu(u)=\frac{dx_1dx_2dx_3}{|x_0|}\end{equation}Thus \begin{equation}I(\lambda)=\int_{x_0>0}(x^2_3-x^2_2-x^2_1)^\lambda_+\phi(x)dx_1dx_2dx_3 \end{equation}$$\phi(x)=\frac{x(u)}{|x_0|}$$ Setting $$x^2_1+x^2_2=v\hspace{1em}x^2_3=u$$
$$v=ut$$ we have \begin{equation}I(\lambda)=-\frac{1}{4}\int
du\,u^{\lambda+\frac{1}{2}}\int^1_0(1-t)^\lambda\psi(u,tu)dt\end{equation}where
\begin{equation}\psi(u,tu)=\int^{4\pi}_0\phi(\sqrt{ut}\cos\eta,\sqrt{ut}\sin\eta,\sqrt u)d\eta\end{equation}\pagebreak

Since $\phi(x)$ vanishes outside a bounded set the integral $I(\lambda)$ converges
in the usual sense for $\mathrm{Re}\lambda>-1$. For $\mathrm{Re}\lambda<-1$ it is to be
understood in the sense of its regularization (analytic continuation) : \begin{equation}I(\lambda)=\frac{1}{4}\frac{1}{(e^{2\pi i\lambda}-1)(e^{2\pi
i\lambda}+1)}\int^{0_+}_\infty du\,
u^{\lambda+\frac{1}{2}}\int^{1_+}_0(1-t)^\lambda\psi(u,tu)dt\end{equation}We now define
\begin{equation}\Phi(\lambda,u)=-\frac{1}{4}\ \frac{1}{(e^{2\pi
i\lambda}-1)}\int^{1_+}_0(1-t)^\lambda\psi(u,tu)dt\end{equation}Just as the generalized
function $(x^\lambda_+,\phi)\ \Phi(\lambda,u)$ is regular for all $\lambda$ except
at
$$\lambda=-1,-2,-3,\ldots$$ where it has simple poles.

On the other hand at regular points of $\Phi(\lambda,u)$ the integral \begin{equation}I(\lambda)=-\frac{1}{(e^{2\pi i\lambda}+1)}\int^{0+}_\infty du\,
u^{\lambda+\frac{1}{2}}\Phi(\lambda,u)\end{equation}also have poles at
$$\lambda=-\frac{3}{2},-\frac{5}{2},\cdots$$ which are once again simple poles.
The analytic function $I(\lambda)$ can be written as \begin{equation}I(\lambda)=\sum^\infty_{\eta=0}\frac{\frac{1}{n!}\left[\frac{\partial^n}{\partial u^n}\Phi(\lambda,u)\right]_{u=0}}{\lambda+\frac{3}{2}+n}+E(\lambda) \end{equation}where
$E(\lambda)$ is an entire function. Thus
$$\mathrm{Res}[I(\lambda)]_{\lambda=-\frac{3}{2}}=\Phi\left(-\frac{3}{2},0\right)=-\frac{1}{2}\psi(0,0)$$
Now $$\psi(0,0)=\int^{4\pi}_0\phi(0,0,0)d\eta=4\pi x(e)$$ Thus \begin{equation}\mathrm{Res}[I(\lambda)]_{\lambda=-\frac{3}{2}}=-2\pi x(e)\end{equation}We shall calculate the same
thing in another way by writing
$$I(\lambda)=\int^{\frac{\pi}{2}}_0\sin^{2\lambda+1}\theta[F(\theta)\sin\theta]$$
We now write $\sin\theta=\theta[1-u(\theta)]$ where
$$\displaystyle u(0)=\left[\frac{du}{d\theta}\right]_{\theta=0}=0$$
\begin{tabbing}where\hspace{1em}\=\kill
Thus\>$\displaystyle I(\lambda)=\int^{\frac{\pi}{2}}_0\theta^{2\lambda+1}G(\theta)d\theta$\`(87a)\\[1.5ex]
where\>$G(\theta)=F(\theta)\sin\theta[1-u(\theta)]^{2\lambda+1}$\`(87b) \end{tabbing} Since
$G(\theta)$ has a compact support and is regular at $\theta=0$ we have\cite{img-ges} for
$\mathrm{Re}\lambda>-\frac{n}{2}-1$ \setcounter{equation}{87}\begin{equation}I(\lambda)=\int^{\frac{\pi}{2}}_0\theta^{2\lambda+1}\left[G(\theta)-\sum^{n-1}_{r=0}\frac{G^{(r)}(0)\theta^r}{r!}\right]+\sum_r\frac{G^{(r)}(0)}{r!(2\lambda+2+r)}
\end{equation}Hence \begin{equation}\mathrm{Res}[I(\lambda)]_{\lambda=-\frac{3}{2}}=\frac{1}{2}G'(0)=\frac{1}{2}\left\{\frac{d}{d\theta}[\sin\theta
F(\theta)]\right\}_{\theta=0}\end{equation}Equating Eqs.\ (86) and (89) we have \begin{equation}x(e)=-\frac{1}{4\pi}\left\{\frac{d}{d\theta}[\sin\theta F(\theta)]\right\}_{\theta=0} \end{equation}Combining Eqs.\ (76) and (90) we immediately obtain
\begin{equation}
x(e)=\frac{2}{\pi^2}\int^\infty_0ds\int^1_0d\tau
\mathrm{Tr}(T^{\left(\frac{\tau}{2},s\right)}_x)\times s\mathrm{Re}\tanh\pi\left(s+\frac{i\tau}{2}\right)
+\frac{2}{\pi^2}\int^\infty_{\frac{1}{2}}dk\left(k-\frac{1}{2}\right)\mathrm{Tr}(T^{k+}_x+T^{k-}_x)
\end{equation}Replacing $x(e)$ by $x^\dagger_1x_2(e)$ and using Eq.\ (20) we have
\vspace{.5ex}\begin{tabbing}\hspace{.4in}\=\kill
\>$\displaystyle\int\overline{x_1(u)}x_2(u)d\mu(u)=\frac{2}{\pi^2}\int^\infty_0ds\int^1_0d\tau
\mathrm{Tr}[T^{\left(\frac{\tau}{2},s\right)^\dagger}_{x_1}T^{\left(\frac{\tau}{2},s\right)}_{x_2}]s\mathrm{Re}\tanh\pi\left(s+\frac{i\tau}{2}\right)$\\[2ex]
\`${}+\displaystyle\frac{2}{\pi^2}\int^\infty_{\frac{1}{2}}dk\left(k-\frac{1}{2}\right)\mathrm{Tr}(T^{k+^\dagger}_{x_1}T^{k+}_{x_2}+T^{k-^\dagger}_{x_1}T^{k-}_{x_2})$\hspace{2em}(92)
\end{tabbing} This is the analogue of the Plancherel formula for the ordinary Fourier
transform.

\section{Acknowledgement}
The author would like to thank the Director, Saha
Institute of Nuclear Physics for providing working facilities and Prof.\ P. S.
Majumdar, Theory Group for helpful discussions.


\begin{thebibliography}{99}\bibitem{img-man} I. M. Gel'fand and M. A. Naimark, {\it I. M. Gel'fand
-- Collected Papers} (Springer-Verlag, Berlin, 1988), Vol II pp.\ 41, 182.
\bibitem{njav-auk} N. Ja. Vilenkin and A. U. Klimyk, {\it Representations of Lie Groups and
Special Functions} (Kluwer Academic, Boston, 1991) Vol I, Chap.\ 6, p.\
298.
\bibitem{img-ges} I. M.\ Gel'fand and G. E.\ Shilov, {\it Generalized Functions} Vol I, p.\
253.
\bibitem{vb-1947} V. Bargmann, Ann, Math.\ {\bfseries48}, 568 (1947).
\bibitem{lp} L. Pukanszky, Math.\ Annalen {\bfseries 156}, 96 (1964).
\bibitem{rh-jaw} R. Herb and J. A.\ Wolf, Compositio Mathematica, {\bfseries 57}(3) 271, (1986).
\bibitem{md-mv} M. Duflo and M.\ Vergne, {\it La formula de Plancherel des groupes de Lie
semi-simples riels.} Advanced Studies in Pure Math.\ {\bfseries 14}, 289 (1988).
\bibitem{db-1997} Subrata Bal, K. V.\ Shajesh and Debabrata  Basu J. Math.\ Phys. {\bfseries38}, 3209
(1997).
\bibitem{vg-1961} V. Bargmann, Commun.\ Pure Appl.\ Math.\ {\bfseries 14}, 187 (1961); {\bfseries 20}, 1
(1967).
\bibitem{ies} I. E.\ Segal, Ill. J.\ Math.\ {\bfseries 6}, 500 (1962).
\bibitem{img-mig-iip} I. M.\ Gel'fand, M. I.\ Graev I. I.\ Pyatetsk\"{u}-Shapiro, {\it Representation
theory and Automorphic Functions}, (Saunders, Philadelphia,  1969). In this reference
$2k-1$ is a positive integer.
\end{thebibliography}
\end{document}